\documentclass[american,aps,pra,superscriptaddress,twocolumn]{revtex4-2}

\usepackage[T1]{fontenc}
\usepackage[sc]{mathpazo}
\usepackage{amsmath}
\usepackage{amssymb}
\usepackage{enumerate}
\usepackage{amsthm}
\usepackage{color}
\usepackage{graphicx} 

\usepackage{hyperref}
\hypersetup{pdfpagemode=UseNone}

\newcommand{\e}{{\rm e}}   

\newcommand{\bra}[1]{\langle#1|}

\newcommand{\ket}[1]{|#1\rangle}

\newcommand{\braket}[1]{\langle#1\rangle}

\begin{document}

\title{Solvable dilation model of time-dependent $\cal PT$-symmetric systems}

 \author{Minyi Huang}
 \email{11335001@zju.edu.cn}
 \email{hmyzd2011@126.com}
  \affiliation{Interdisciplinary Center of Quantum Information, State Key Laboratory of Modern Optical Instrumentation, and Zhejiang Province Key Laboratory of Quantum Technology and Device, Department of Physics, Zhejiang University, Hangzhou 310027, China}
 \affiliation{Department of Mathematical Sciences, Zhejiang Sci-Tech University, Hangzhou 310018, PR~China}

\author{Ray-Kuang Lee}
 \email{rklee@ee.nthu.edu.tw}
\affiliation{Institute of Photonics Technologies, National Tsing Hua University, Hsinchu 300, Taiwan}
\affiliation{Department of Physics, National Tsing Hua University, Hsinchu 300, Taiwan}
\affiliation{Physics Division, National Center for Theoretical Sciences, Taipei 10617, Taiwan}
\affiliation{Center for Quantum Technology, Hsinchu 30013, Taiwan}

\author{Qing-hai Wang}
\email{qhwang@nus.edu.sg}
\affiliation{Department of Physics, National University of Singapore, Singapore 117551, Singapore}

\author{Guo-Qiang Zhang}
\email{zhangguoqiang@csrc.ac.cn}
\affiliation{Interdisciplinary Center of Quantum Information, State Key Laboratory of Modern Optical Instrumentation, and Zhejiang Province Key Laboratory of Quantum Technology and Device, Department of Physics, Zhejiang University, Hangzhou 310027, China}


 \author{Junde Wu}
 \email{wjd@zju.edu.cn}
\affiliation{School of Mathematical Sciences, Zhejiang University, Hangzhou 310027, PR~China}

\begin{abstract}
The dilation method is a practical way to experimentally simulate non-Hermitian, especially $\cal PT$-symmetric quantum systems. However, the time-dependent dilation problem cannot be explicitly solved in general. In this paper, we present a simple yet non-trivial exactly solvable dilation problem for two-dimensional time-dependent $\cal PT$-symmetric Hamiltonians. Our system is initially set in the unbroken $\cal PT$-symmetric phase, then goes across the so-called exceptional point, and ends in  the broken $\cal PT$-symmetric phase. For this system, the dilated Hamiltonian and the evolution of $\cal PT$-symmetric system are analytically worked out.
 By investigating the large time behaviors, we give an effective method to choose and adjust the dilation parameters.
 Our result also shows that the exceptional points do not have much physical relevance in a \textit{time-dependent} system.
\end{abstract}

\maketitle

\section{Introduction}
\label{sec:intro}

In recent years, researchers have witnessed a growing interest in discussing non-Hermitian systems, especially in the field of dynamics and topology \cite{ashida2020non}. Lots of work has been done and many intriguing properties of non-Hermitian systems are revealed and discussed.
The related topics, such as skin effect, attracts much increasing attentions \cite{PhysRevLett.121.086803,PhysRevLett.121.136802,alvarez2018non,ozdemir2019parity,PhysRevLett.123.170401,PhysRevLett.123.246801,PhysRevResearch.1.023013}.

As one of the most important classes of non-Hermitian systems, $\cal PT$-symmetric systems are of great interests
both theoretically and experimentally. The systematic studies on such systems began in 1998, with Bender and his colleagues' discussion on the reality of the eigenvalues of $\cal PT$-symmetric Hamiltonians \cite{bender1998real}. Since then, much work has been done to investigate $\cal PT$-symmetric quantum systems, among which Mostafazadeh generalized $\cal PT$-symmetric theory to pseudo-Hermitian theory~\cite{mostafazadeh2002Pseudo1,mostafazadeh2002Pseudo2,mostafazadeh2002Pseudo3,mostafazadeh2010pseudo}. %
Recently, there are also discussions on anti-$\cal PT$-symmetric systems \cite{wu2015parity,wang2016optical}.

In general, $\cal PT$-symmetric systems are non-Hermitian and it is possible to use large Hermitian systems to simulate such non-Hermitian systems. The simulation of $\cal PT$-symmetric systems is tightly related to the mathematical concept of operator
dilation. In 2008, G\"{u}nther and Samsonov showed that a special two-dimensional unbroken $\cal PT$-symmetric Hamiltonian can be
dilated and their results were experimentally realized \cite{gunther2008naimark,tang2016experimental}. Later, the result was generalized
to any finite-dimensional case \cite{PhysRevLett.119.190401,huang2018embedding}. As for the broken $\cal PT$-symmetry, there are also different approaches. One way is utilizing weak measurement,
which can be viewed as an approximation paradigm
\cite{PhysRevLett.123.080404}; while the other way is simulating the time-dependent broken $\cal
PT$-symmetric systems with the time-dependent Hermitian systems \cite{wu2019observation,zhang2019time}.
In fact, time-dependent $\cal PT$-symmetric systems are important research issues of their own rights, e.g., the Floquet theory such as and many other features of these systems \cite{Faria2006time,Mostafazadeh2007time,Znojil2008time,Fring2016nonHermitian,Fring2016Unitary,Maamache2017pseudo,
Maamache2017invariant,Mostafazadeh2020time,Fring2020Spectrally,Fring2021Exactly,Fring2021infinite,Fring2021introduction,Fring2022time}. In particular, the work using Dyson maps by Fring and collaborators implies that exceptional points (EP) do not play an essential role in such time-dependent systems \cite{Fring2020Spectrally,Fring2021Exactly,Fring2021infinite,Fring2021introduction,Fring2022time}.
The discussion of time-dependent dilation gives an important approach to investigating the topology and dynamics of non-Hermitian systems. However,  the problem is that usually the time evolution operator and the dilated Hamiltonian cannot be analytically worked out, owing to the fact that the Hamiltonian at different time cannot be diagonalized in the same eigenstates \cite{wu2019observation,liu2021dynamically}.

In this paper, we discuss a solvable example for the time-dependent dilation problem. All the relevant matrix operators are worked out explicitly. Our model shows that the exceptional points have no physical significance in a \textit{time-dependent} system as the dynamics throughout smoothly evolves.

The paper is organised as follows. In Sec.~\ref{sec:prelim}, we briefly review the elements of dilation. In Sec.~\ref{sec:solvable}, we discuss the time-dependent dilation problem and give a solvable model. In Sec.~\ref{sec:special}, we present the detailed results of a special case. We make some discussions in Sec.~\ref{sec:discussion}, and conclude our results in In Sec.~\ref{sec:conclusion}.

\section{The concept of dilation}
\label{sec:prelim}

In this section, we briefly recap the dilation method described in \cite{gunther2008naimark,wu2019observation,zhang2019time}. Consider an $n$-dimensional, time-dependent non-Hermitian Hamiltonian $H(t)$, it governs an evolution by the Schr\"odinger equation,
\begin{equation}
	i\dot{\psi}(t)=H(t)\psi(t),
\label{eqn:schr1}
\end{equation}
where the overhead dot denotes the time derivative. The unit with $\hbar=1$ is adopted. For simplicity, we may suppress the time variable. To simulate such a system in experiments, we dilate the state into $2n$-dimensions, i.e., $\Psi:=\begin{bmatrix}\psi\\\tau\psi\end{bmatrix}$, where $\tau$ is an ancillary matrix to be specified later. The dilated vector $\Psi$ evolves under a \textit{Hermitian} Hamiltonian
\begin{equation}
	i\dot{\Psi}(t) = \mathbb{H}(t)\Psi(t).
\label{eqn:schr2}	
\end{equation}
Here, $\mathbb{H}:=\begin{bmatrix}h_1&h_2\\h_2^\dag & h_4\end{bmatrix}$ with $h_1=h_1^\dag$ and $h_4=h_4^\dag$.
Equations (\ref{eqn:schr1}) and (\ref{eqn:schr2}) yield the following conditions,
\begin{eqnarray}
h_1+h_2\tau&=&H,\label{h21}\\
h_2^\dag+h_4\tau&=&i\dot{\tau}+\tau H.
\label{tau}
\end{eqnarray}
It follows from Eq.~(\ref{tau}) that
\begin{eqnarray}
h_2=-i\dot{\tau}^\dag+H^\dag\tau^\dag-\tau^\dag h_4.\label{H41}
\end{eqnarray}
By substituting Eq. (\ref{H41}) into Eq. (\ref{h21}), we have
\begin{equation}
h_1=H+i\dot{\tau}^\dag\tau-H^\dag\tau^\dag\tau+\tau^\dag h_4\tau.\label{H2}
\end{equation}
Thus, the dilated Hamiltonian $\mathbb H$ is determined by $h_4$ and $\tau$. Apparently, $h_4$ can be an arbitrary $n\times n$ Hermitian matrix and one only needs to find $\tau$.
By the Hermiticity of $h_1$, we have
\begin{equation}
i\frac{d}{dt}(\tau^\dag\tau)=H^\dag(\mathbb{1}+\tau^\dag\tau) - (\mathbb{1}+\tau^\dag\tau)H.
\label{H4}
\end{equation}
If we denote
\begin{equation}
	\eta(t):=(\mathbb{1}+\tau^\dag\tau),
	\label{eqn:tau}	
\end{equation}
then
\begin{equation} i\dot{\eta}=H^\dag\eta-\eta H.
\label{eta}
\end{equation}
By construction, $\eta$ is positive definite.
If $\eta$ happens to be time-independent, Eq.~(\ref{eta}) indicates that the non-Hermitian Hamiltonian is $\cal PT$-symmetric, $H=H_{\cal PT}$ with unbroken $\cal PT$ symmetry,
\begin{equation}
H_{\cal PT}^\dag\eta=\eta H_{\cal PT}.
\label{e1}
\end{equation}
Equation (\ref{e1}) is often called pseudo-Hermiticity in the literature. On the other hand, when $\eta$ is positive definite, the condition of $\eta$-pseudo-Hermiticity is equivalent to unbroken $\cal PT$-symmetry in finite dimensional spaces \cite{huang2018embedding}. We will use the term of $\cal PT$ symmetry throughout this paper.
In general, an arbitrary non-Hermitian $H(t)$ is a combination of an (unbroken) $\cal PT$-symmetric Hamiltonian $H_{\cal PT}$ and a gauge term \cite{zhang2019time,PTQM},
\begin{equation}
	H = H_{\cal PT} - \frac{i}{2} \eta^{-1} \dot{\eta}.
\end{equation}

Usually, $\eta$ is called the metric operator. The key to dilate a non-Hermitian system is to find a metric operator such that Eq.~(\ref{eta}) holds. An important observation is that the matrix $\tau$ exists in Eq.~(\ref{eqn:tau}) if and only if $(\eta-\mathbb{1})$ is semi-positive definite. Or equivalently, all the eigenvalues of the Hermitian matrix $(\eta-\mathbb{1})$ are nonnegative. In this case, we can always write the solution of Eq.~(\ref{eqn:tau}) in the polar decomposition as $\tau=U\sqrt{\eta-\mathbb{1}}$, where $U$ is an arbitrary unitary matrix. Different choice of $U$ will lead to different but equivalent dilation. For simplicity, we choose a Hermitian $\tau$ with $U=\mathbb{1}$. Obtaining $\tau$, one can further construct the large Hermitian Hamiltonian $\mathbb{H}$. Note that the Hermitian Hamiltonian $\mathbb{H}$ is not determined because $h_4$ is an arbitrary $n\times n$ Hermitian matrix. One simple way to specify $h_4$ is to take it as the Hermitian part of $H$, $h_4 = \frac{1}{2} (H+H^\dag)$. Another way is to follow Ref.~\cite{wu2019observation} and demand $h_4=[H+(i\dot{\tau}+\tau H)\tau ]\eta^{-1}$.


The above formalism allows to simulate, namely effectively realize non-Hermitian systems using larger Hermitian systems.


To find $\eta$, let us take
\[\eta(t)=\zeta^\dag(t)\zeta(t),\]
where $\zeta(t)$ is a matrix satisfying the following differential equation \cite{zhang2019time},
\begin{equation}
i\dot{\zeta}^\dag(t)=H^\dag(t)\zeta^\dag(t).
\label{H}
\end{equation}
The solutions to Eq.~(\ref{H}) can be easily constructed from the solutions to the dual Schr\"odinger equation whose Hamiltonian is $H^\dag(t)$. Note that the initial value of $\zeta(0)$ is arbitrary as long as all the moduli of its eigenvalues are not smaller than $1$. Different choices of $\zeta(0)$ lead to different but equivalent $\eta(t)$ and the dilation Hamiltonian $\mathbb{H}(t)$ \cite{zhang2019time}.

In general, a closed form of the solution to Eq.~(\ref{H}) is hard to find. In the next section, we discuss  a simple but non-trivial two-dimensional model for which its dilation problem can be solved exactly.


\section{A Solvable Model}
\label{sec:solvable}

In this section, we illustrate the general ideas using a concrete example. We start with a $2\times2$ time-dependent non-Hermitian Hamiltonian $H_\omega(t)$ and solve the Schr\"odinger equation governed by it. Our goal is to obtain the dilated Hermitian Hamiltonian $\mathbb H$ as explicit as possible. Equations (\ref{H41}) and (\ref{H2}) show that $\mathbb H$ is determined by $\tau$. The key step to obtain $\tau$ is to find the metric operator $\eta$ determined by Eq.~(\ref{eta}). The discussion near the end of Sec.~\ref{sec:prelim} shows that $\eta$ can be constructed by the solutions to the dual Schr\"odinger equation. Finally, by taking a square-root of $(\eta-\mathbb{1})$, we get $\tau$.

The $2\times2$ time-dependent Hamiltonian is as follows,
\begin{equation}
H_\omega(t)=
\begin{bmatrix}
	E+i\omega t & 1\\
	1& E-i\omega t
\end{bmatrix},\label{13}
\end{equation}
where $E$ and $\omega$ are real parameters. The parity operator is chosen to be the first Pauli matrix and the time reversal operator to be the complex conjugation (or Hermite conjugation since the matrix is symmetric),
\begin{equation}
	\mathcal{P}=\sigma_x, \qquad \mathcal{T}=*~\mathrm{or}~\dag.
\end{equation}
One can verify that $H_\omega(t)$ is $\cal PT$-symmetric, that is,
\begin{equation}
	H_\omega(t){\cal PT}={\cal PT}H_\omega(t).
	\label{eqn:PT}
\end{equation}	
Such an example can be used to discuss the Jarzynski equality \cite{deffner2015jarzynski} and the time-dependent $\cal PT$-symmetric quantum mechanics \cite{zhang2019time,PTQM}. The instantaneous eigenvalues are
\begin{equation}
	\lambda(t) = E \pm \sqrt{1-\left(\omega t\right)^2}.
\end{equation}
When $\omega t < 1$, the ${\cal PT}$ symmetry is unbroken, and both eigenvalues are real. At the exceptional point (EP), $\omega t = 1$, the Hamiltonian is not diagonalizable. When $\omega t >1$, the eigenvalues are complex. Thus, the ${\cal PT}$ symmetry is broken for $\omega t \geq 1$. One may expect that some critical phenomena happen at the EP. On the contrary, as we will see later, this is not the case. The dynamics evolves smoothly even when $\omega t$ crosses the EP.

We divide the process of solving dilation into several subsections. In Subsec.~\ref{subsec:schr}, we solve the Schr\"{o}dinger equation governed by $H_\omega(t)$,
\begin{equation}
	i\dot{\psi}(t)=H_\omega(t)\psi(t).
	\label{eqn:Homega}
\end{equation}	
In Subsec.~\ref{sec:metric}, we solve the dual Schr\"{o}dinger equation to construct $\eta$. In Subsec.~\ref{subsec:tau}, we give the form of $\tau$ by taking square-root of $(\eta-\mathbb{1})$. Thus, by arbitrarily choosing a Hermitian matrix $h_4$, one can obtain $\mathbb H$ and the dilation problem is solved. In Subsec.~\ref{subsec:largetime}, we discuss some large time behavior which dictates when a given dilation may fail.

\subsection{Solutions to the Sch\"{o}dinger equation (\ref{eqn:Homega})}
\label{subsec:schr}
We write the solution to Eq.~(\ref{eqn:Homega}) in the following component form,
\[
\psi(t)=
\begin{bmatrix}
	x_\uparrow(t)\\
	x_\downarrow(t)
\end{bmatrix}.
\]
Now Eq.~(\ref{eqn:Homega}) gives two combined equations,
\begin{eqnarray}
	i\dot{x}_\uparrow (t)& = & (E+i\omega t)x_\uparrow (t) +x_\downarrow (t), 	\label{eqn:xupdown1}\\
	i\dot{x}_\downarrow (t)& =& x_\uparrow(t)+(E-i\omega t)x_\downarrow (t). 		\label{eqn:xupdown}
\end{eqnarray}
By substituting Eq.~(\ref{eqn:xupdown1}) into Eq.~(\ref{eqn:xupdown}) and eliminating $x_\downarrow(t)$, we get a second order differential equation,
\begin{equation}
	\ddot{x}_\uparrow (t)+2iE\dot{x}_\uparrow (t)+\left[1-E^2-\omega\left(1+wt^2\right)\right]x_\uparrow(t)=0.
	\label{eqn:xupddot}
\end{equation}
By changing variables,
\begin{equation}
z :=\omega t^2\quad \mathrm{and} \quad w(z) := \sqrt{t}\,\e^{iEt}x_\uparrow(t),
\end{equation}
we obtain a Whittaker equation,
\begin{equation}
w''(z)+\left(-\frac{1}{4}+\frac{1-\omega}{4\omega z}+\frac{3}{16z^2}\right)w(z)=0.
\end{equation}
The general solution can be represented by the Whittaker functions,
\begin{equation}
w(z)=C_0 W_{\kappa,\mu}(z)+C_1W_{-\kappa,\mu}(\e^{i\pi}z)\label{W1}
\end{equation}
with
\[
	\kappa =-\frac{1}{4}+\frac{1}{4\omega},\quad \mu=\frac{1}{4}.
\]
Here we follow the notations in Ref.~\cite{DLMF}. In terms of the original variables, one has
\begin{eqnarray}
\nonumber x_\uparrow (t) = C_0\frac{\e^{-iEt}}{\sqrt{t}}W_{\kappa,\mu}(\omega t^2)+C_1\frac{\e^{-iEt}}{\sqrt{t}}W_{-\kappa,\mu}(-\omega t^2).\\
\label{W2}
\end{eqnarray}
For simplicity, let us define two linearly independent solutions as
\begin{equation}
	\begin{aligned}
		&x_\uparrow^{(0)}(t) := \frac{\e^{-iEt}}{\sqrt{t}} W_{\kappa,\mu}(\omega t^2),\\
		&x_\uparrow^{(1)}(t) := \frac{\e^{-iEt}}{\sqrt{t}} W_{-\kappa,\mu}(-\omega t^2).
	\end{aligned}
\label{eqn:xupper}
\end{equation}
Note that there is no singularity as $t\to 0$ because $W_{\pm\kappa,\mu}(\pm\omega t^2) \propto \sqrt{t}$ for small $t$ \cite[(13.14.18)]{DLMF}. In principle, the corresponding lower components $x_\downarrow^{(i)}$ can be solved similarly by eliminating $x_\uparrow^{(i)}$ from Eq.~(\ref{eqn:xupdown1}). Moreover, the coefficients in $x_\downarrow^{(i)}$ are determined by the corresponding $x_\uparrow^{(i)}$. After a lengthy calculation (see Appendix \ref{sec:lower}), compact results are found,
\begin{equation}
	\begin{aligned}
		&x_\downarrow^{(0)}(t) = -2i\e^{-iEt}\sqrt{\frac{\omega}{t}} W_{\kappa',\mu}(\omega t^2),\\
		&x_\downarrow^{(1)}(t) = \frac{\e^{-iEt}}{2\sqrt{\omega t}} W_{-\kappa',\mu}(-\omega t^2),
	\end{aligned}
\label{x1}
\end{equation}
where $\kappa'=\frac{1}{4}+\frac{1}{4\omega}.$ Note that all four Whittaker functions are smooth near the EP, implying that nothing special happens. Furthermore, since the Wronskian (${\cal W}$) of the Whittaker functions is a constant~ \cite{DLMF}, i.e.,
$${\cal W}\{W_{\kappa,\mu}(z), W_{-\kappa,\mu}(\e^{\pi i}z)\}=\e^{-\kappa\pi i},$$
these two solutions will never coalesce.

It is also known that for some special values of parameters, the Whittaker function can be truncated to Hermite polynomials~\cite{DLMF},
\begin{equation}
W_{\frac{1}{4}+\frac{n}{2},\frac{1}{4}}(z)=\frac{\e^{-z/2}}{2^n}z^{1/4}H_n(\sqrt{z}).
\end{equation}
Therefore, when $\frac{1}{2\omega}$ is an integer, one of the solutions reduces to elementary functions.

\subsection{The metric operator}
\label{sec:metric}
To determine the metric operator $\eta(t)$, we need to solve Eq.~(\ref{H}), where $H^\dag(t)$ is now given by $H_\omega^\dag(t)$. The column vectors of $\zeta^\dag$ are just the solutions to the "dual Schr\"odinger equation",
\begin{equation}
i\dot{\phi}(t)=H_\omega^\dag(t) \phi(t).
\label{dual}
\end{equation}
On the other hand, one can derive from the $\mathcal{PT}$ symmetry condition in Eq.~(\ref{eqn:PT}) that
\[
\sigma_x H_\omega =H_\omega^\dag\sigma_x.
\]
Let $\psi$ be a solution to the Schr\"odinger equation (\ref{eqn:Homega}), then
\begin{equation}
H_\omega^\dag \sigma_x\psi=\sigma_x H_\omega \psi =i \frac{d}{dt}(\sigma_x\psi).
\end{equation}
That is, $\phi=\sigma_x\psi$ is the solution to the dual Schr\"odinger equation (\ref{dual}). If we define
\begin{eqnarray*}
	y^{(0)}:=\sigma_x x^{(1)}\quad \mathrm{and}\quad
	y^{(1)}:=\sigma_x x^{(0)},
\end{eqnarray*}
then the explicit forms of $y^{(i)}$ are
\begin{eqnarray}
y^{(0)}(t)&=&\frac{\e^{-iEt}}{\sqrt{t}}
\begin{bmatrix}
\frac{1}{2\sqrt{\omega}}W_{-\kappa',\mu}(-\omega t^2)\\
W_{-\kappa,\mu}(-\omega t^2)
\end{bmatrix},\\
y^{(1)}(t)&=&\frac{\e^{-iEt}}{\sqrt{t}}
\begin{bmatrix}
-2i\sqrt{\omega}W_{\kappa',\mu}(\omega t^2)\\
W_{\kappa,\mu}(\omega t^2)
\end{bmatrix}.
\end{eqnarray}
The two column vectors of $\zeta^\dag$ are both linear combinations of $y^{(0)}$ and $y^{(1)}$. To determine $\zeta^\dag$, one should determine four linear combination coefficients.
For simplicity, we consider a solution to Eq.~(\ref{H}) with only two parameters $D_0$ and $D_1$,
\begin{eqnarray}
	\zeta^\dag(t)=\begin{bmatrix}
		D_0 y_\uparrow^{(0)}(t) &D_1 y_\uparrow^{(1)}(t)\\
		D_0 y_\downarrow^{(0)}(t) &D_1 y_\downarrow^{(1)}(t)
	\end{bmatrix}.
\end{eqnarray}
With this choice, we have
\begin{eqnarray}
\eta=\zeta^\dag\zeta=|D_0|^2\ket{y^{(0)}}\bra{y^{(0)}}+|D_1|^2\ket{y^{(1)}}\bra{y^{(1)}},
\label{eqn:eta}
\end{eqnarray}
where the bras and kets are conventional Dirac notation with $\ket{\cdot}=\bra{\cdot}^\dag$.
From the definition of $\eta$ in Eq.~(\ref{eqn:tau}), $(\eta-\mathbb{1})$ must be semi-positive definite. Thus it requires that all the eigenvalues of $\eta$ are not smaller than one. Otherwise, one cannot find appropriate matrix $\tau$ such that Eq.~(\ref{H4}) holds. In this case, dilation fails.
This imposes constraints on $D_i$. In this model, for any finite time interval, one can always find a set of appropriate $D_i$ such that the eigenvalues of $\eta$ are not smaller than one, that is, the dilation can be valid over any given time interval. For this purpose, note that $\eta=\zeta^\dag\zeta$ has the eigenvalues
\begin{equation}
\lambda_\pm =\frac{l}{2}\pm \sqrt{\frac{l^2}{4}-|D_0D_1|^2\Delta},
\label{eg}
\end{equation}
where we define
\begin{eqnarray*}
	l&:=& |D_0|^2\|y^{(0)}\|^2 + |D_1|^2\|y^{(1)}\|^2, \nonumber\\
	\Delta &:=& \|y^{(0)}\|^2\|y^{(1)}\|^2-\left|\braket{y^{(0)}|y^{(1)}}\right|^2
\end{eqnarray*}
with the conventional notation $\|\cdot\| := \sqrt{\braket{\cdot|\cdot}}$.

There are various ways to choose appropriate $D_i$ for the eigenvalues to be not smaller than one.
Moreover, since $\lambda_+\geqslant\lambda_-$, one only needs to guarantee that $\lambda_-\geqslant 1$.
For example, one may choose $D_0=D_1=D$. According to the Schwartz inequality,
$
\Delta>0.
$
Denote $\tilde{l}:=\|y^{(0)}\|^2+\|y^{(1)}\|^2$, then $\lambda_-\geqslant 1$ is equivalent to
\begin{equation*}
	|D|^2\geqslant\frac{\tilde{l}+\sqrt{\tilde{l}^2-4\Delta}}{2 \Delta}.
\end{equation*}
Note that right-hand-side is a continuous function of $\tilde{l}$ and $\Delta$. Thus, in any finite time interval, it has a maximal value. We may always choose $|D|^2$ to be larger than this maximum to ensure the dilation to be valid.

\subsection{The dilation matrix $\tau$}
\label{subsec:tau}
After solving the metric operator $\eta(t)$, we are ready to find the dilation operator $\tau(t)$ in the dilated state vectors $\Psi$ and Hamiltonian $\mathbb{H}$. As discussed before, we choose a Hermitian $\tau$ for simplicity,
\begin{equation}
	\tau=\sqrt{\eta-\openone}=
	\begin{bmatrix}
		d+c&a-ib\\
		a+ib&d-c\\
	\end{bmatrix}.
\label{eqn:tau1}	
\end{equation}
It can be shown that (see Appendix \ref{sec:Apptau} for more details)
\begin{eqnarray}
a&=&\frac{X}{\sqrt{2}\sqrt{W+\sqrt{W^2-X^2-Y^2-Z^2}}}, \nonumber\\
b&=&\frac{Y}{\sqrt{2}\sqrt{W+\sqrt{W^2-X^2-Y^2-Z^2}}}, \nonumber\\
c&=&\frac{Z}{\sqrt{2}\sqrt{W+\sqrt{W^2-X^2-Y^2-Z^2}}}, \nonumber\\
d&=&\frac{\sqrt{W+\sqrt{W^2-X^2-Y^2-Z^2}}}{\sqrt{2}},
\label{eqn:abcd}
\end{eqnarray}
where $X$, $Y$, $Z$, and $W$ are matrix elements of $\eta-\openone$,
\begin{equation}
	\tau^2=\eta-\openone=
	\begin{bmatrix}
		W+Z&X-iY\\
		X+iY&W-Z\\
	\end{bmatrix}.
\label{eqn:tau2}	
\end{equation}
According to Eq. (\ref{eqn:eta}),
 the explicit forms of the matrix elements of $\tau^2$ are
\begin{eqnarray}
&&X=\frac{1}{2}\left[|D_0|^2\left(y_\uparrow^{(0)}y_\downarrow^{(0)*}+y_\uparrow^{(0)*}y_\downarrow^{(0)}\right)\right. \nonumber\\
&&\qquad\left.+|D_1|^2\left(y_\uparrow^{(1)}y_\downarrow^{(1)*}+y_\uparrow^{(1)*}y_\downarrow^{(1)}\right)\right], \nonumber\\
&&Y=\frac{i}{2}\left[|D_0|^2\left(y_\uparrow^{(0)}y_\downarrow^{(0)*}-y_\uparrow^{(0)*}y_\downarrow^{(0)}\right)\right. \nonumber\\
&&\qquad \left.+|D_1|^2\left(y_\uparrow^{(1)}y_\downarrow^{(1)*}-y_\uparrow^{(1)*}y_\downarrow^{(1)}\right)\right], \nonumber\\
&&Z=\frac{1}{2}\left[|D_0|^2\left(|y_\uparrow^{(0)}|^2-|y_\downarrow^{(0)}|^2\right)\right. \nonumber\\
&&\qquad \left.+|D_1|^2\left(|y_\uparrow^{(1)}|^2-|y_\downarrow^{(1)}|^2\right)\right], \nonumber\\
&&W=\frac{1}{2}\left(|D_0|^2\|y^{(0)}\|^2+|D_1|^2\|y^{(1)}\|^2\right)-1,
\label{eqn:xyz}
\end{eqnarray}
and
\begin{eqnarray}
	&&W^2-X^2-Y^2-Z^2 \nonumber\\
	&&= 1-\left(|D_0|^2\|y^{(0)}\|^2+|D_1|^2\|y^{(1)}\|^2\right) \nonumber\\
	&&+|D_0D_1|^2\left(\|y^{(0)}\|^2\|y^{(1)}\|^2-|\braket{y^{(0)}|y^{(1)}}|^2\right).
	\label{eqn:w}
\end{eqnarray}
With a proper choice of $h_4$, one can get the explicit dilated Hamiltonian $\mathbb{H}$. Regardless of the choice on $h_4$, Eqs.~(\ref{H41}) and (\ref{H2}), as well as the more or less complicated form of $\eta$, make the final form of $\mathbb{H}$ too long to be presented here.


\subsection{Large time behavior}
\label{subsec:largetime}

The above discussions only apply to the situation of a finite time interval. When time $t$ tends to be infinitely large, the simulation will always fail in this model. We may see this by studying the large time behaviors.

Applying formulas in Ref.~\cite{DLMF}, one can derive the large time behaviors for the eigenvalues of $\eta$ as $t\to\infty$ (see Appendix \ref{sec:Applarget} for details),
\begin{eqnarray}
	\lambda_+ & \sim & \frac{\sqrt{\omega}}{(\omega t^2)^\frac{1}{2\omega}} |D_0|^2 \e^{\omega t^2}, \label{eqn:lambda+} \\
	\lambda_- & \sim & |D_1|^2 4\omega ^{\frac{3}{2}} (\omega t^2)^\frac{1}{2\omega} \e^{-\omega t^2}. \label{eqn:lambda-}
\end{eqnarray}
The decaying factor $\e^{-\omega t^2}$ in $\lambda_-$ indicates that the eigenvalue will be smaller than one for sufficiently large time, regardless of the choices on $D_i$. That is, all dilation schemes will break down eventually. To quantitatively see when the dilation will fail, we will later discuss for a special case in detail.

\section{The special case with $\omega=\frac{1}{2}$}
\label{sec:special}

In this section, we present the results for a special case with $\omega=\frac{1}{2}$ in details. The Hamiltonian becomes
\begin{equation}
H_\frac{1}{2}(t)=\begin{bmatrix}
	E+\frac{1}{2}ti & 1\\
	1& E-\frac{1}{2}ti
\end{bmatrix}.
\end{equation}
The eigenvalues of $H_\frac{1}{2}(t)$ are $\lambda_\pm=E\pm \sqrt{1-\frac{t^2}{4}}$. We consider a finite time interval $t\in [0,4]$. When $t\in [0,2)$, the eigenvalues $\lambda_\pm$ are real and $\cal PT$-symmetry is unbroken. When $t\in (2,4]$, $\lambda_\pm$ are complex and thus $\cal PT$-symmetry is broken. The EP occurs at $t=2$.

For this case, the solution in Eq.~(\ref{W2}) is
\begin{equation}
	x_\uparrow (t) = C_0 \frac{\e^{-iEt}}{\sqrt{t}}W_{\frac{1}{4},\frac{1}{4}}\left(\frac{1}{2} t^2\right) + C_1 \frac{\e^{-iEt}}{\sqrt{t}}W_{-\frac{1}{4},\frac{1}{4}}\left(-\frac{1}{2} t^2\right).
\end{equation}
Note that
$W_{\frac{1}{4},\frac{1}{4}}\left(\frac{1}{2} t^2\right) \propto \sqrt{t}\,\e^{-\frac{t^2}{4}}$ and
$W_{-\frac{1}{4},\frac{1}{4}}\left(-\frac{1}{2} t^2\right)$ is a linear combination of
$\sqrt{t}\, \e^{-\frac{t^2}{4}}$ and $\sqrt{t}\, \e^{-\frac{t^2}{4}}\mathrm{Erfi}\left(\frac{t}{\sqrt{2}}\right)$ \cite[(13.18.7) and (13.18.16)]{DLMF}, where $\mathrm{Erfi}(z)$ is the imaginary error function (we adopt the definition in Ref.~\cite{DLMF}, \textit{i.e.},  $\mathrm{Erfi}(z):=\int_0^z e^{x^2}dx$). Therefore, for convenience, we choose one solution vector as
\begin{eqnarray}
	&&\ket{x^{(0)}(t)}
	=\e^{-iEt-\frac{t^2}{4}}\begin{bmatrix}
		\alpha
		\\
		\beta
	\end{bmatrix}\label{eqn:A}\\
&& \mathrm{with} \quad \alpha=1 \quad \mathrm{and} \quad \beta=-it \nonumber
\end{eqnarray}
and the other solution vector as
\begin{eqnarray}
	&&\ket{x^{(1)}(t)}=\e^{-iEt-\frac{t^2}{4}}\begin{bmatrix}
		\gamma
		\\
		\delta
	\end{bmatrix}, \label{eqn:D}\\
	\mathrm{where}\qquad 	&& \gamma=-i\sqrt{2}\, \mathrm{Erfi}[\frac{t}{\sqrt{2}}], \nonumber\\
	&& \delta=\e^{\frac{t^2}{2}}-\sqrt{2}\,t \,\mathrm{Erfi}[\frac{t}{\sqrt{2}}].\nonumber
\end{eqnarray}

To determine the metric operator $\eta(t)$, we take two independent solutions to the dual Schr\"{o}dinger equation as
\begin{eqnarray}
	&&\ket{y^{(0)}(t)}=\e^{-iEt-\frac{t^2}{4}}\begin{bmatrix}
		\delta
		\\
		\gamma
	\end{bmatrix}, \\
	&&\ket{y^{(1)}(t)}
	=\e^{-iEt-\frac{t^2}{4}}\begin{bmatrix}
		\beta
		\\
		\alpha
	\end{bmatrix},
\end{eqnarray}
where $\alpha, \beta, \gamma,$ and $\delta$ are defined in Eqs.~(\ref{eqn:A}) and (\ref{eqn:D}).

It can be verified that
\begin{eqnarray*}
	\ket{x^{(0)}(0)}=\ket{y^{(0)}(0)}=\ket{0} &=& \begin{bmatrix}
		1\\
		0
	\end{bmatrix}, \\
	\ket{x^{(1)}(0)}=\ket{y^{(1)}(0)}=\ket{1}&=&\begin{bmatrix}
		0\\
		1
	\end{bmatrix}.
\end{eqnarray*}
This justifies our labeling on $\ket{x}$ and $\ket{y}$. In addition, as time increase, $\ket{y^{(0)}}$ and $\ket{x^{(1)}}$ tend to be infinitely large while $\ket{y^{(1)}}$ and $\ket{x^{(0)}}$ tend to vanish.

\begin{figure}
	\centering
	\includegraphics[width=80 mm]{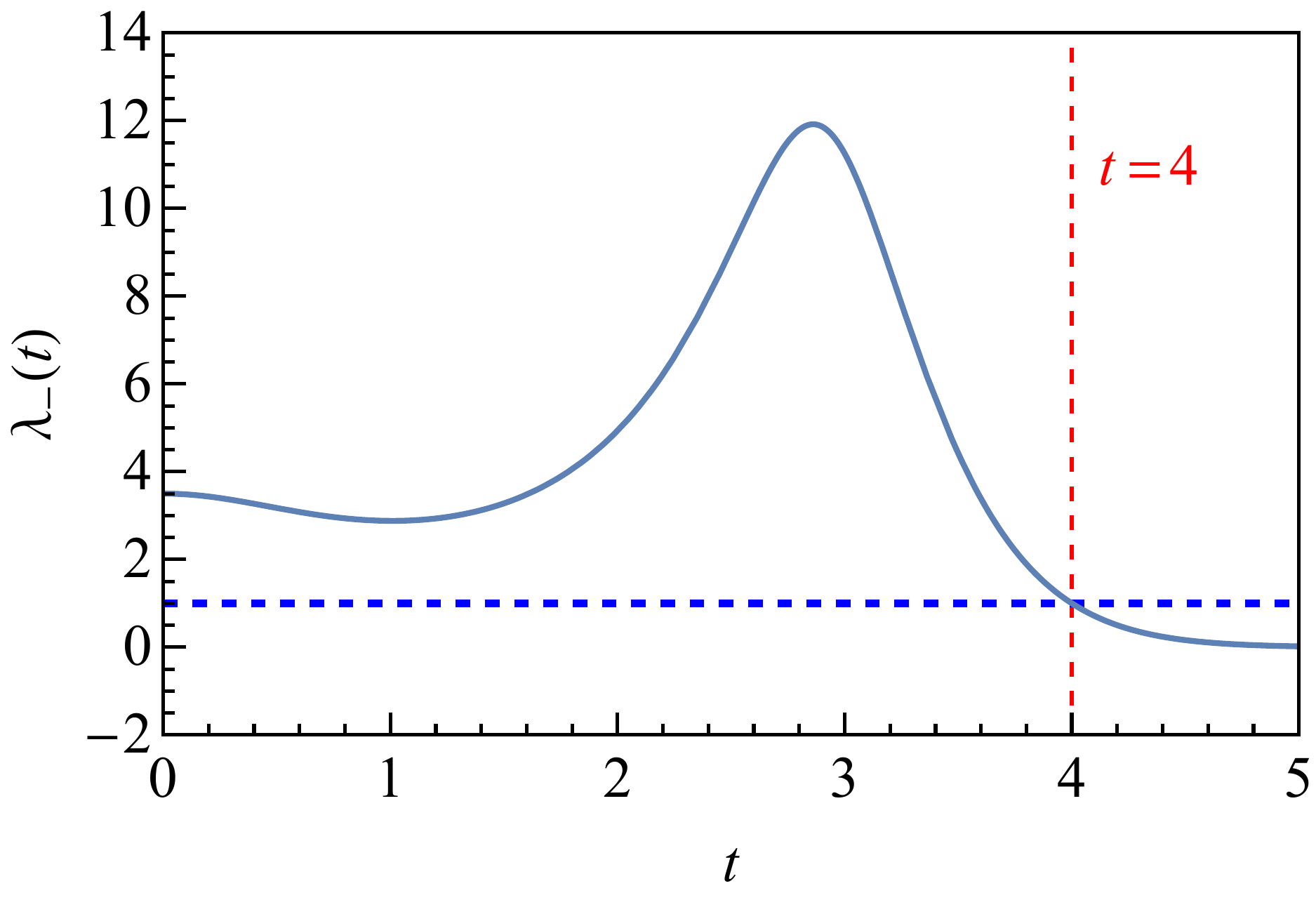}
	\includegraphics[width=80 mm]{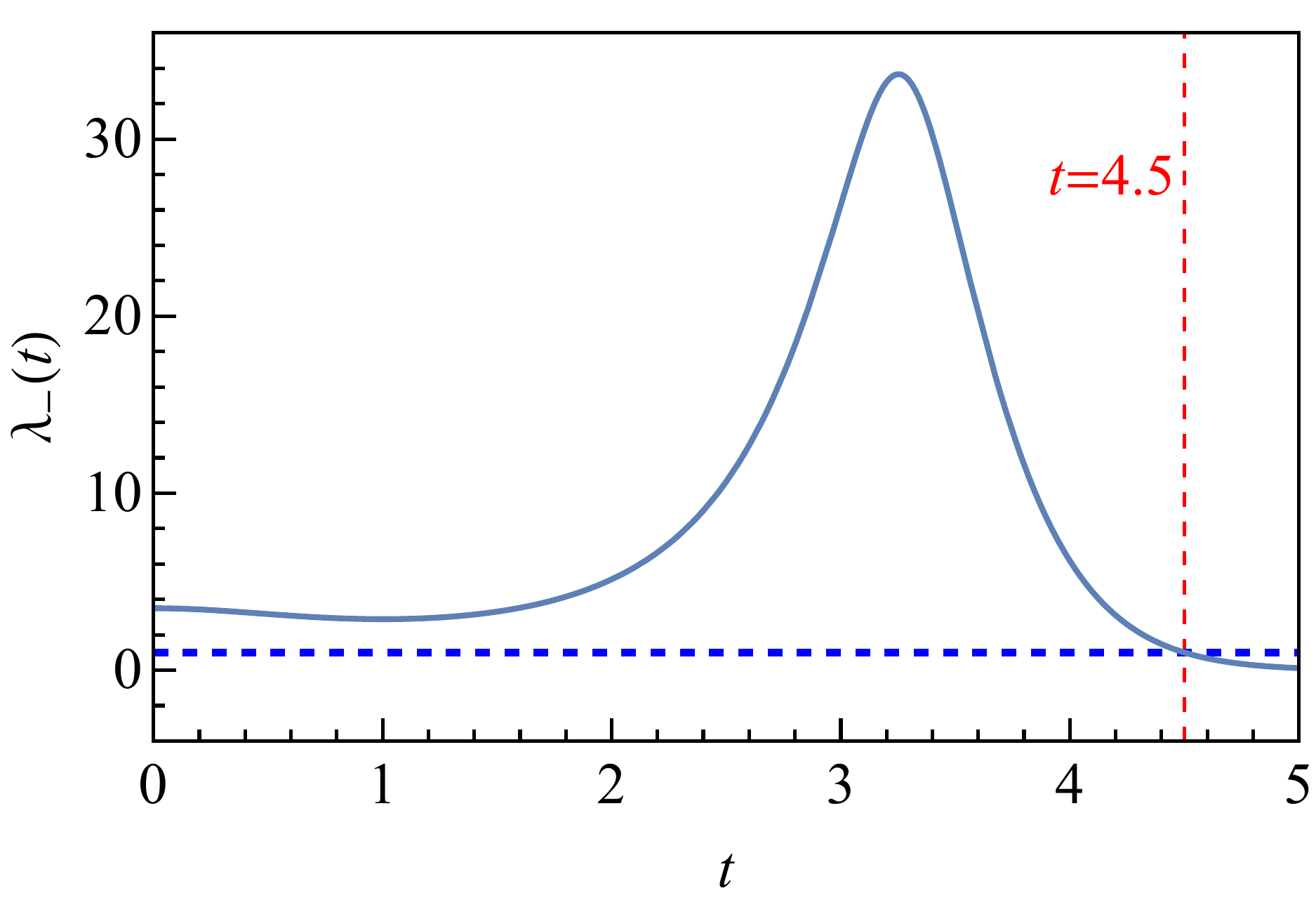}
	\caption{The smaller eigenvalue $\lambda_-$ of the metric $\eta$ as a function of time for $t\in [0,5]$. Here we choose $D_0^2=3.5$ for both panels. In the upper panel, $D_1^2=238$. The dilation is valid up to $t\approx4.0001$. In the lower panel, $D_1^2=1474$, and the dilation is valid up to $t\approx 4.5$. The solid lines are the smaller eigenvalue $\lambda_-$ of the metric operator $\eta$. The (blue) dashed lines are the constant $1$. The dilations are valid when they are above the horizontal (blue) dashed lines. The vertical (red) dashed lines mark the time limits when the dilations fail. Note that both eigenvalues are smooth at the exceptional point $t=2$. }
	\label{lm2}
\end{figure}

Now the problem is to find appropriate parameters $D_i$ which give a successful dilation in a finite time interval. As shown in Subsec.~\ref{sec:metric}, this reduces to finding $D_i$ such that $\lambda_-\geqslant1$ over the target time domain.
 The large time behaviors often help us to choose the parameters $D_i$. To see this, we deduce from Eq.~(\ref{eg}) that $\lambda_-\geqslant1$ is equivalent to
\begin{equation}
2\leqslant l\leqslant 1+ |D_0|^2 |D_1|^2, \label{d1}
\end{equation}
where we substitute the result of $\Delta=1$ in this model.
Now suppose that we want to obtain $D_i$ which gives a successful dilation over the time interval $[0,4]$.
Note that for large $t$, e.g., $t=4$ in the present case, $\|y^{(1)}\|$ is small. Therefore, Eq.~(\ref{d1}) may be approximated as
\begin{eqnarray}
&&|D_0|^2 \geqslant \mathrm{max}\frac{2}{\|y^{(0)}(t)\|^2}\approx 3.43,\label{c1}\\
&&|D_1|^2 \geqslant \mathrm{max}\|y^{(0)}(t)\|^2\approx 237.80.\label{c2}
\end{eqnarray}
Thus, a possible way to choose parameters is to take $D_i$ satisfying Eqs.~(\ref{c1}) and (\ref{c2}) and verify whether these parameters really ensure the dilation. For example, we can take $D_0^2=3.5$ and $D_1^2=238$. The smaller eigenvalue of the metric operator,  $\lambda_-(t)$ is plotted in the top panel of Fig.~\ref{lm2}. The dilation fails around $4.0001$.

For this model, $y^{(0)}$ dominates for large time. When $|D_0|$ is fixed, in order to extend the valid time for the dilation to $t_0$, we simply need to choose
\begin{equation}
	|D_1|^2 \geqslant \|y^{(0)}(t_0)\|^2.
	\label{bt}
\end{equation}
In the lower panel of Fig.~\ref{lm2}, we plot a case which is valid up to $t=4.5$.

\begin{figure}
	\centering
	\includegraphics[width=80 mm]{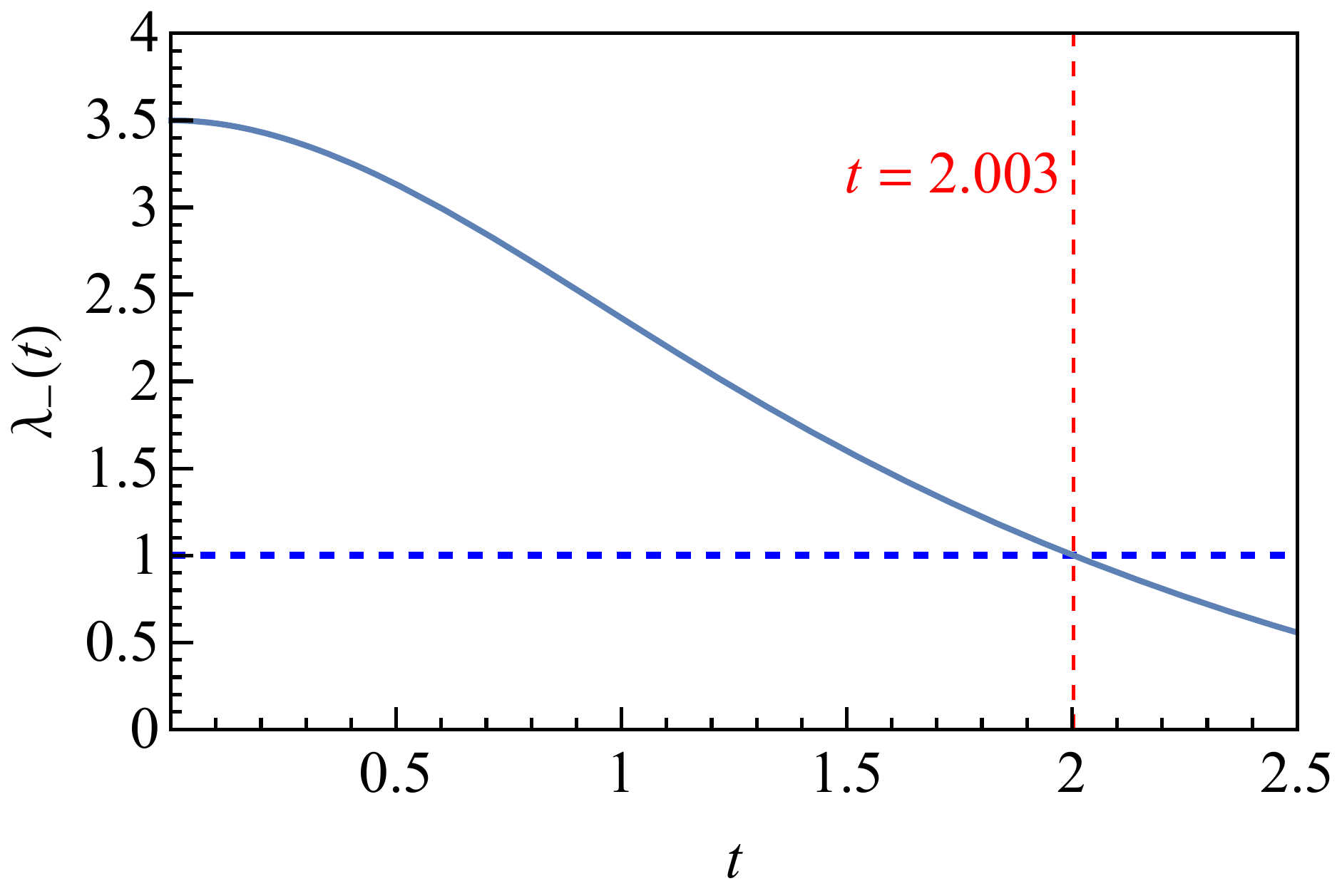}
	\includegraphics[width=80 mm]{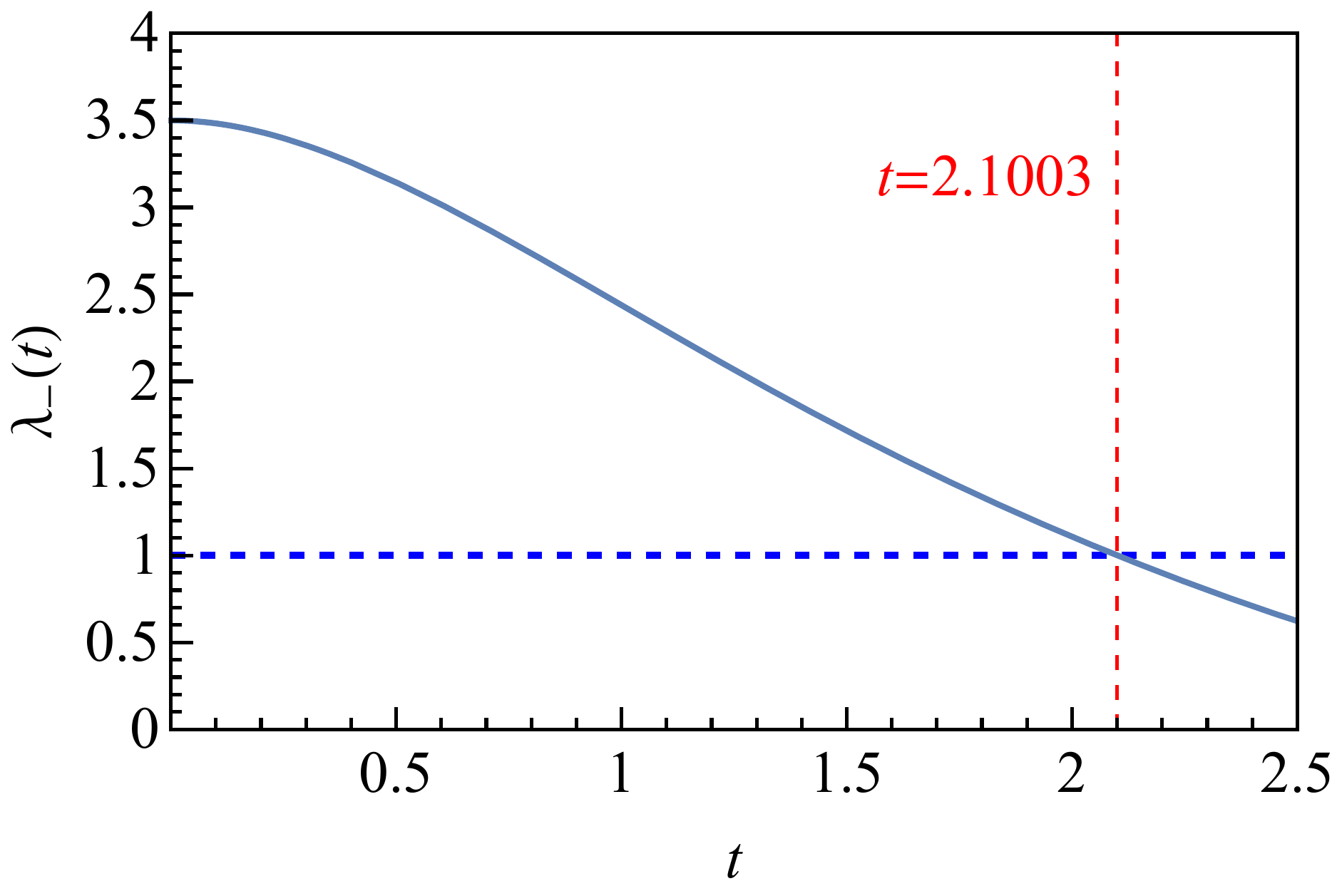}
	\caption{The smaller eigenvalue $\lambda_-$ of $\eta$ as a function of time for $t\in[0,2.5]$. Here we choose $D_0^2=3.5$ for both panels. For the upper panel, $D_1^2=4.13$ and the dilation is valid up to $t\approx2.003$. For the lower panel, $D_1^2=4.634$ and the dilation is valid up to $t\approx 2.1003$. The solid lines are the smaller eigenvalue of $\eta$. The (blue) dashed lines are constant $1$. The dilations are valid when the solid line are above the horizontal (blue) dashed lines. The vertical (red) dashed lines mark the time when the dilations fail. Note again that nothing dramatic happens at the exceptional point $t=2$.}
	\label{lm3}
\end{figure}

Note that Eq.~(\ref{bt}) gives a very good estimation of the breakdown time for $t_0=4$ or lager. This is because $y^{(0)}$ is exponentially large for large $t$. However, the asymptotic analysis may need to be fine-tuned for not so large dilation interval. For example, if we consider the time interval $[0, 2.1]$ and $D_0^2=3.5$, one may guess that $|D_1|^2\geqslant|y^{(0)}(2.1)|^2\approx 4.129$ according to Eq.~(\ref{bt}). However, the dilation actually breaks down earlier at $t\approx2.003$ for $D_1^2=4.13$. In such a situation, we must use the full expression in Eq.~(\ref{d1}) to make a better estimation, which is equivalent to
\begin{equation}
|D_1|^2\geqslant\frac{|D_0|^2\|y^{(0)}\|^2-1}{|D_0|^2-\|y^{(1)}\|^2}.
\label{bt2}
\end{equation}
For the dilation to be valid up to $t=2.1$, we must take  $|D_1|^2 \geqslant  \frac{3.5\|y^{(0)}(2.1)\|^2-1}{3.5-\|y^{(1)}(2.1)\|^2}\approx 4.633$. The actual break down time is $t\approx 2.1003$ with $D_1^2=4.634$. The smaller eigenvalue of the metric operator of both choices are plotted in Fig.~\ref{lm3}. Once again, both functions are smooth around the exceptional point at $t=2$.

\section{Discussions}
\label{sec:discussion}

The key to specify a metric operator is to determine the parameters $D_0$ and $D_1$. They can either take the same or different values. An advantage of taking different values is to improve the efficiency of dilation. According to Eq.~(\ref{eqn:schr2}), for the state $\Psi=\begin{bmatrix}\psi\\\tau\psi\end{bmatrix}$, the $\cal PT$-symmetric system is simulated for the upper components. Hence the dilation efficiency can be characterized by $\frac{\braket{\psi|\psi}}{\braket{\psi|\mathbb{1}+\tau^\dag\tau|\psi}} = \frac{\braket{\psi|\psi}}{\braket{\psi|\eta|\psi}}$. In our discussion of time interval $[0,4]$, $|D_1|^2\geqslant 238$. If we choose $|D_0|=|D_1|$, the dilation efficiency for any state can be estimated by $\frac{1}{238^2}$. However, if we take $|D_0|^2=3.5$, then for state $\begin{bmatrix}\ket{0}\\ \tau(0) \ket{0}\end{bmatrix}$,  its dilation efficiency can be characterized by $\frac{1}{3.5^2}> \frac{1}{238^2}$. Thus the dilation efficiency is improved.

The EP plays a central role in many studies on non-Hermitian Hamiltonians. A critical phenomenon is expected at the exceptional point in a \textit{time-independent} system because the energy eigenstates coalesce and the Hamiltonian becomes non-diagonalizable. However, in a \textit{time-dependent} system like ours, the instantaneous eigenstates are not solutions to the Schr\"odinger equation. The critical behavior of the eigenstates does not directly translate to the dynamical states of the system. Therefore, one should not expect anything special to happen at the exceptional point. As shown in our model, the exceptional point is as normal as any other point within the dilation interval.

As shown in Sec.~\ref{sec:solvable}, for the given Hamiltonian in Eq.~(\ref{13}) and any finite time interval, one can always find appropriate $\eta$ (and $\tau$) such that the dilation is valid. However, as $t$ tends to infinity, the dilation will eventually break down. Note that this breakdown time can be arbitrarily postponed by different dilation parameters $D_i$. Thus, it cannot be an \textit{intrinsic} critical point of the original non-Hermitian system. Rather, such a breakdown is only a limitation of our dilation technique.

It might be interesting from the mathematical point of view to see what happens to the dilated Hamiltonian $\mathbb H$ after the breakdown time. If we keep the ancillary matrix as $\tau=U\sqrt{\eta-\mathbb{1}}$ with an arbitrary unitary matrix $U$, then the dilated Hamiltonian $\mathbb H$ defined by Eqs.~(\ref{H41}) and (\ref{H2}) is no longer Hermitian. To see this, note that $\sqrt{\eta-\mathbb{1}}$ is not Hermitian for $\lambda_-<1$. Therefore,
\[
\tau^\dag\tau+\mathbb{1}=(\sqrt{\eta-\mathbb{1}})^\dag(\sqrt{\eta-\mathbb{1}})+{\mathbb{1}}\neq \eta.
\]	
Plugging the above inequality into Eq.~(\ref{H2}), in general we have that
\begin{equation}
	 h_1-h_1^\dag \neq  i\dot{\eta} -H^\dag \eta + \eta H = 0.
	 \label{dh4}
\end{equation}
Further note that the non-Hermiticity of $\mathbb{H}$ cannot be saved from any choice of $h_4$.

\section{Conclusion}
\label{sec:conclusion}

In summary, a two-dimensional solvable model for  \textit{time-dependent} $\cal PT$-symmetric systems is shown to have an explicit scheme to dilate into a four-dimensional Hermitian system.
Furthermore, by investigating the large time behaviors, we give an effective method to choose and adjust the dilation parameters.
A good estimation of the breakdown time for the dilation is also derived.
As the dilated Hermitian Hamiltonians play an important role in the simulation of $\cal PT$-symmetric systems, our results may shed new light on the study of \textit{time-dependent} $\cal PT$-symmetric systems.


\section*{Acknowledgement}
This work is partially supported by the National Natural Science Foundation of China (Grants Nos. 11901526, 11971140, 12031004 and 61877054), the China Postdoctoral Science Foundation (Grant No. 2020M680074), the Natural Science Foundation of Zhejiang Province (Grant No. LY22A010010), the Science Foundation of Zhejiang Sci-Tech University (Grant No. 19062117-Y) and the Fundamental Research Foundation for the Central Universities (Project No.K20210337).

\appendix
\section{The lower component of the solution to the Schr\"{o}dinger equation}
\label{sec:lower}

In this Appendix, we show how to derive the lower component of the solution to the Schr\"odinger equation in Eq.~(\ref{x1}). Using Eq.~(\ref{eqn:xupdown}), one easily expresses Eq.~(\ref{eqn:xupdown1}) as a
second order differential equation for $x_\downarrow$, which is very similar to that of $x_\uparrow$ in (\ref{eqn:xupddot}),
\begin{equation}
	\ddot{x}_\downarrow+2iE\dot{x}_\downarrow+\left[1-E^2+\omega\left(1-\omega t^2\right)\right]x_\downarrow=0.
	\label{eqn:xdownddot}
\end{equation}
The only difference in Eqs.~(\ref{eqn:xupddot}) and (\ref{eqn:xdownddot}) is the sign of $\omega$. Thus, one can immediately read off the general solution for $x_\downarrow$ is the linear combination of two Whittaker functions,
\begin{equation}
	x_\downarrow (t) = C_3\frac{\e^{-iEt}}{\sqrt{t}}W_{\kappa',\mu}(\omega t^2)+C_4\frac{\e^{-iEt}}{\sqrt{t}}W_{-\kappa',\mu}(-\omega t^2),
	\label{eqn:xdown}
\end{equation}
where $\kappa':=\frac{1}{4}+\frac{1}{4\omega}.$ One may guess that
\begin{eqnarray}
	x_\downarrow^{(0)}(\omega,t) &=& C_3 \frac{e^{-iEt}}{\sqrt{t}}W_{\kappa',\frac{1}{4}}(\omega t^2),\\
	x_\downarrow^{(1)}(\omega,t) &=& C_4 \frac{e^{-iEt}}{\sqrt{t}}W_{-\kappa',\frac{1}{4}}(-\omega t^2).
\end{eqnarray}
To determine the correct constants $C_3$ and $C_4$, let us plug Eq.~(\ref{eqn:xupper}) into Eq.~(\ref{eqn:xupdown1}),
\begin{equation}
	x_\downarrow= i\dot{x}_\uparrow-(E+i\omega t)x_\uparrow. \label{A5}
\end{equation}
Applying the identity \cite[(13.15.25)]{DLMF} with $n=1$,
\[
\frac{d}{dz}
\left[\e^{-\frac{1}{2}z}z^{\mu-\frac{1}{2}} W_{\kappa,\mu}(z)\right] = - \e^{-\frac{1}{2}z}z^{\mu-1}W_{\kappa+\frac{1}{2},\mu-\frac{1}{2}}(z)
\]
and recognizing that \cite[(13.14.31)]{DLMF},
\[
W_{\kappa,\mu}(z) = W_{\kappa,-\mu}(z),
\]
we get from Eq.~(\ref{A5}) that
\begin{equation}
x_\downarrow^{(0)} = -2i\omega^{1\over 2}\frac{\e^{-iEt}}{\sqrt{t}}W_{\kappa',\frac{1}{4}}(\omega t^2).
\end{equation}
That is,
\[C_3=-2i\sqrt{\omega}.\]

Similarly, by applying \cite[(13.15.22)]{DLMF} with $n=1$,
\begin{eqnarray*}
&&\frac{d}{dz}\left[\e^{\frac{1}{2}z}z^{\mu -\frac{1}{2}}W_{\kappa,\mu}(z)\right] \nonumber\\
&=& -\left(\frac{1}{2}-\kappa-\mu\right) \e^{\frac{1}{2}z}z^{\mu-1}W_{\kappa-\frac{1}{2},\mu-\frac{1}{2}}(z),
\end{eqnarray*}
we obtain,
\begin{equation*}
x_\downarrow^{(1)} = \frac{1}{2\sqrt{\omega}} \frac{\e^{-iEt}}{\sqrt{t}}W_{-\kappa',\frac{1}{4}}(-\omega t^2).
\end{equation*}
Namely, $C_4=\frac{1}{2\sqrt{\omega}}$. Putting them together, the explicit forms of the two independent solutions to the Schr\"odinger equation are
\begin{eqnarray}
x^{(0)}(t)=\frac{\e^{-iEt}}{\sqrt{t}}
\begin{bmatrix}
W_{-\frac{1}{4}+\frac{1}{4\omega},\frac{1}{4}}(\omega t^2)\\
-2i\sqrt{\omega}W_{\frac{1}{4}+\frac{1}{4\omega},\frac{1}{4}}(\omega t^2)
\end{bmatrix},\label{eqn:x0}\\
x^{(1)}(t)=\frac{\e^{-iEt}}{\sqrt{t}}
\begin{bmatrix}
W_{\frac{1}{4}-\frac{1}{4\omega},\frac{1}{4}}(-\omega t^2)\\
\frac{1}{2\sqrt{\omega}}W_{-\frac{1}{4}-\frac{1}{4\omega},\frac{1}{4}}(-\omega t^2)
\end{bmatrix}, \label{eqn:x1}
\end{eqnarray}
where $-\omega t^2 = \e^{i\pi} \omega t^2$ for the branch-cut of the Whittaker function.

\section{Large time behaviours}
\label{sec:Applarget}
In this Appendix, we derive the large time behaviors of the dilated system. Applying the identity \cite[(13.14.21)]{DLMF},
\begin{equation}
W_{\kappa,\mu}\sim \e^{-\frac{1}{2}z}z^\kappa,\quad \mathrm{as} \quad z\to \infty, \quad |\arg z| < \frac{3}{2} \pi -\delta
\end{equation}
to the solution of the Schr\"odinger equation in Eqs.~(\ref{eqn:x0}) and (\ref{eqn:x1}), we get the large time behaviors,
\begin{eqnarray}
	x^{(0)}
	&\sim &
	\e^{-iEt-\frac{1}{2}\omega t^2} \omega^{-\frac{1}{4}+\frac{1}{4\omega}}t^{\frac{1}{2\omega}-1}
	\begin{bmatrix}
		1\\
		-2i\omega t
	\end{bmatrix},
	\label{xt1}\\
	x^{(1)}
	&\sim &
	\e^{-iEt+\frac{1}{2}\omega t^2} (\e^{i\pi}\omega)^{\frac{1}{4}-\frac{1}{4\omega}}t^{-\frac{1}{2\omega}}
	\begin{bmatrix}
		1\\
		\frac{1}{2i\omega t}
	\end{bmatrix}.\label{xt2}
\end{eqnarray}
Accordingly, the large time behaviors of the solutions to the ``dual equation'' are
\begin{eqnarray}
y^{(0)}
&\sim & \e^{-iEt+\frac{1}{2}\omega t^2} (-\omega)^{\frac{1}{4}-\frac{1}{4\omega}}t^{-\frac{1}{2\omega}}
\begin{bmatrix}
\frac{1}{2i\omega t}\\
1
\end{bmatrix},\label{yt1}\\
y^{(1)}
&\sim & \e^{-iEt-\frac{1}{2}\omega t^2} \omega^{-\frac{1}{4}+\frac{1}{4\omega}}t^{\frac{1}{2\omega}-1}
\begin{bmatrix}
-2i\omega t\\
1
\end{bmatrix}.\label{yt2}
\end{eqnarray}

The large time behaviors of $y^{(i)}$ implies that one of the eigenvalues of $\eta$ will tend to vanish. This implies that the dilation will fail when $t$ is sufficiently large. In fact, using Eqs.~(\ref{yt1}) and (\ref{yt2}),  we see that
\begin{eqnarray*}
y_{\uparrow\downarrow}^{(0)}\rightarrow \infty &,\qquad& y_{\uparrow\downarrow}^{(1)}\rightarrow 0,\label{yd2}\\
y_\downarrow^{(0)}\sim 2i\omega t y_\uparrow^{(0)}&,\qquad& y_\downarrow^{(1)}\sim -\frac{1}{2i\omega t}y_\uparrow^{(1)}.\label{yd}
\end{eqnarray*}
Thus one can further obtain
\begin{equation}
	l= D_0^2\|y^{(0)}\|^2+D_1^2\|y^{(1)}\|^2
	\sim 4\omega^2t^2|D_0y_\uparrow^{(0)}|^2,
\label{B4}
\end{equation}
and
\begin{equation}
\Delta = y_\uparrow^{(0)}y_\downarrow^{(1)}-y_\uparrow^{(1)}y_\downarrow^{(0)}
\sim  -2i\omega t y_\uparrow^{(0)}y_\uparrow^{(1)}.\label{B5}
\end{equation}
Plugging Eqs. (\ref{B4}) and (\ref{B5}) into Eq.~(\ref{eg}), we get the large time behaviors of the two eigenvalues of $\eta$ in Eqs.~(\ref{eqn:lambda+}) and (\ref{eqn:lambda-}).

\section{The form of $\tau (t)$}
\label{sec:Apptau}
In the discussion of dilation problem, $\tau=\sqrt{\eta-\openone}$ is semi-positive definite. Direct calculation from Eqs.~(\ref{eqn:tau1}) and (\ref{eqn:tau2}) leads to
\begin{eqnarray}
X&=&2ad,\label{x}\\
Y&=&2bd,\label{y}\\
Z&=&2cd.\label{z}\\
W&=&a^2+b^2+c^2+d^2,
\label{w}
\end{eqnarray}
By substituting Eqs.~(\ref{x})-(\ref{z}) into Eq.~(\ref{w}), we have
\begin{equation}
d^2=\frac{1}{2}\left(W\pm\sqrt{W^2-X^2-Y^2-Z^2}\right).
\end{equation}
Since $\tau$ is semi-positive definite, its trace and determinant are both nonnegative. These mean that $d\geqslant0$ and $d^2\geqslant a^2+b^2+c^2$. Combining with Eq.~(\ref{w}), we know $d^2\geqslant\frac{1}{2}W\geqslant0$.
Thus we have Eq.~(\ref{eqn:abcd}) in the main text.

\end{document}